1  **Natural recovery of genetic diversity by gene flow in reforested areas of the**
2  **endemic Canary Island pine,** *Pinus canariensis*
3

4  Miguel Navascués, Brent C. Emerson
5
6  Centre for Ecology, Evolution and Conservation, School of Biological Sciences,
7  University of East Anglia, Norwich NR4 7TJ, United Kingdom
8
9  Telephone:  (44)  01603  592237.  Fax:  (44)  01603  592250.  E-mail:
10  b.emerson@uea.ac.uk
11
12  Correspondence address: Brent Emerson, School of Biological Sciences, University of
13  East Anglia, Norwich NR4 7TJ, UK.
14

15  **Abstract**

16  The endemic pine, *Pinus canariensis*, forms one of the main forest ecosystems in the
17  Canary Islands. In this archipelago, pine forest is a mosaic of natural stands (remnants
18  of past forest overexploitation) and artificial stands planted from the 1940's. The
19  genetic makeup of the artificially regenerated forest is of some concern. The use of
20  reproductive material with uncontrolled origin or from a reduced number of parental
21  trees may produce stands ill adapted to local conditions or unable to adapt in response
22  to environmental change. The genetic diversity within a transect of reforested stands
23  connecting two natural forest fragments has been studied with nuclear and chloroplast
24  microsatellites. Little genetic differentiation and similar levels of genetic diversity to
25  the surrounding natural stands was found for nuclear markers. However, chloroplast microsatellites presented lower haplotype diversity in reforested stands, and this may be a consequence of the lower effective population size of the chloroplast genome, meaning chloroplast markers have a higher sensitivity to bottlenecks. Understory natural regeneration within the reforestation was also analysed to study gene flow from natural forest into artificial stands. Estimates of immigration rate into artificially regenerated forest were high (0.68–0.75), producing a significant increase of genetic diversity (both in chloroplast and nuclear microsatellites) which indicates the capacity for genetic recovery for *P. canariensis* reforestations surrounded by larger natural stands.

**1. Introduction**

Silvicultural practices can potentially affect the genetic structure of forests and this problem is of concern for forest geneticists (see Lefèvre, 2004, for a review). In particular, artificial regeneration of forest (plantations or sowing) may produce a loss of genetic diversity through the processes of unplanned selection and genetic drift (Ledig, 1992). Detection of a reduction in genetic diversity with neutral markers because of artificial selection is unlikely because only a few genes might be involved (Lefèvre, 2004) but the quantification of genetic drift to assess potential losses of adaptive variation can be done from neutral molecular markers as it is a process that affects the whole genome (Glaubitz et al., 2003b).

49  Bottleneck effects may be present within forest regeneration practises because of the
50  use of a limited number of seed trees and the unevenness of the number of seeds
51  collected or produced per tree (Glaubitz et al., 2003a; Burgarella, 2004).

52

53  An additional concern for artificial populations is the extent of genetic exchange with
54  natural populations. Potential negative effects of gene flow from artificial plantations
55  into natural forests ('genetic pollution') have been considered (Lenormand, 2002) and
56  have gained recent public awareness with the use of transgenic crops (including tree
57  species, DiFazio et al., 2004). Gene flow from natural population into seed orchards
58  (i.e. seed production plantations for reforestation) is also considered a negative effect
59  ('pollen contamination') because it reduces the genetic gain obtained from the
60  breeding programs and may increase maladaptation (El-Kassaby, 2000). However,
61  under other circumstances, gene exchange between artificial and natural populations
62  can also be positive. For small declining populations gene flow from surrounding
63  reforestation may increase their effective population size ('demographic rescue',
64  Lefèvre, 2004). Also, gene flow from natural forest into reforested stands may be a
65  natural way to recover the genetic variation lost in the reforestation process. This
66  could potentially create an admixed landrace well adapted to local conditions, and
67  with high evolutionary potential to respond to environmental changes.

68

69  *Pinus canariensis* is endemic to the western Canary Islands, where it forms one of the
70  main forest ecosystems of the archipelago. Its area of distribution has been diminished
71  by five centuries of overexploitation (decreasing from covering 25% of the territory to
72  now covering 12%). A great reforestation effort has been made since the 1940's,
73  resulting in a mosaic of reforested and natural pine forest (del Arco Aguilar et al.,

1992). A high genetic differentiation has been found among populations of *P. canariensis* (Gómez et al., 2003) with some endangered marginal populations presenting particular genetic characteristics (Vaxevanidou et al., 2006). In the present study we analyse the genetic diversity and immigration rates within a transect of reforested stands connecting two natural forest fragments in Tenerife. Hence, we are assessing the effects of the reforestation process on the genetic integrity of the artificial stands and their potential to naturally recover to the levels of diversity of the surrounding natural forest.

**2. Materials and methods**

*2.1 Plant material and sampling design*

El Teide mountain (Tenerife, Spain) is crowned by a pine forest belt (the Corona Forestal Natural Park). A proportion of these pine woods are artificial stands planted from the 1940's to repair the past overexploitation (Parsons, 1981; del Arco Aguilar et al., 1992). We have chosen as our study site an area where there is still a small interruption of the forest with only a narrow strip of planted trees connecting two natural forest fragments (figure 1). Artificial *Pinus canariensis* stands were planted in Fasnia between 1956 and 1965 and in Arico between 1981 and 1985. The current vegetation cover for those areas is 30–60% for Fasnia and < 30% for Arico (del Arco Aguilar et al., 1992). The provenance of the plants is unknown but most probably seed were collected from different locations in Tenerife island (Climent et al., 1996). Nine sites were sampled in these plantations (sites 2–6 in Arico and 7–10 in Fasnia, figure 1); in each of the nine sampling stations 10 planted adult trees and 10 understory naturally regenerated seedlings were sampled (except at site eight where 11 planted

99  trees and 10 seedlings were sampled). From the natural forest in Arico (site 1) and

100 Güímar (site 11) we have sampled 50 and 47 mature trees respectively. Both old trees

101 and seedlings were selected randomly and with an average separation of 10 m among

102 them. From each individual needles were collected and preserved in silica gel in the

103 summer of 2002.

104

105 *2.2 Molecular markers*

106 Genomic DNA was purified using a CTAB protocol based on the Doyle and Doyle

107 (1987) method. Samples were genotyped for eight chloroplast microsatellites (Pt1254,

108 Pt15169, Pt26081, Pt30204, Pt36480, Pt71936, Pt87268 and Pt110048; Vendramin et

109 al., 1996) and eight nuclear microsatellites (SPAC 11.5, SPAC 11.8 and SPAG 7.14,

110 Soranzo et al., 1998; PtTX3116 and PtTX4001, Auckland et al., 2002; ssrPt_ctg4363,

111 ssrPt_ctg4698 and ssrPt_ctg7731, Chagné et al., 2004). PCR amplifications were

112 performed in a Perkin-Elmer 9700 thermal cycler (details in Navascués, 2005). PCR

113 products were sized using an ABI PRISM 3700 DNA Analyzer (Applied Biosystems).

114

115 *2.3 Genetic diversity indices*

116 For chloroplast microsatellites, total number of haplotypes ($n_h$, direct count of

117 haplotypes which are scored as unique combinations of alleles for all cpSSR loci),

118 effective number of haplotypes ($n_e$, Nielsen et al., 2003), unbiased haplotype diversity

119 ($H_e$, Nei, 1978) and average genetic distances among individuals ($D^2_{sh}$, Goldstein et

120 al., 1995) were calculated for each sample:

121 $$n_e = (n-1)^2 \bigg/ \sum_{h=1}^{n_h} p_h^2 (n+1)(n-2) + 3 - n \qquad [1]$$

122 $$H_e = \frac{n}{n-1}\left(1 - \sum_{h=1}^{n_h} p_h^2\right) \qquad [2]$$

123 $$D_{sh}^2 = \frac{2}{n(n-1)} \frac{1}{L} \sum_{i=1}^{n} \sum_{j=i+1}^{n} \left( \sum_{k=1}^{L} |a_{ik} - a_{jk}| \right)^2 \quad [3]$$

124 where $n$ is the number of individuals in the sample, $p_h$ is the relative frequency of the

125 $h^{th}$ haplotype, $n_h$ is the number of different haplotypes in the sample, $L$ is the number

126 of loci, $a_{ik}$ is the size (number of repeats) of the allele for the $i^{th}$ individual and at the

127 $k^{th}$ locus, and $a_{jk}$ is the size of the allele for the $j^{th}$ individual and at the $k^{th}$ locus.

128

129 For nuclear microsatellites the arithmetic mean among loci was calculated for the

130 following indices: number of alleles ($A$, direct count), effective number of alleles ($A_e$,

131 Nielsen et al., 2003), and unbiased gene diversity ($H_e$, Nei, 1978):

132 $$A_e = (2n-1)^2 \bigg/ \sum_{i=1}^{A} p_i^2 (2n+1)(2n-2) + 3 - 2n \quad [4]$$

133 $$H_e = \frac{2n}{2n-1} \left( 1 - \sum_{i=1}^{A} p_i^2 \right) \quad [5]$$

134 where $n$ is the number of individuals in the sample (i.e. sample size), $p_i$ is the relative

135 frequency of the $i^{th}$ allele in the sample and $A$ is the total number of different alleles in

136 the sample.

137

138 Significant differences in the genetic diversity indices between pairs of samples were

139 tested by a Monte Carlo approach (see, for instance, Glaubitz et al., 2003a for similar

140 analysis) programmed in Fortran language (source code available from the authors on

141 request). The null distribution (both samples hold the same genetic diversity) of

142 differences for each index was constructed by resampling (with replacement)

143 individuals among samples. The result is a mixture of individuals from both samples

144 distributed in two new samples (with sizes equal to the original sizes). The genetic

145 diversity parameters of both samples and their differences are computed (samples are compared always in the same order and sign of differences is kept). This process is repeated 10 000 times to build the null distribution of the genetic diversity differences expected for two samples (with the specific sizes of the empirical samples) drawn from populations with the same genetic diversity (i.e. the same population). The empirical difference value is then compared with the null distribution. Since we are interested in testing for the reduction of genetic diversity (i.e. null hypothesis, $h_0$: diversity in natural forest ≤ diversity in artificial stands) we perform a one-tailed test. The *p*-value for rejecting the null hypothesis is estimated as the proportion of iterations with differences greater than or equal to the observed difference value. This process is used to test chloroplast haplotype diversity, individual nuclear locus diversity and average (among loci) nuclear diversity. Two types of comparisons were made, 1) samples from the natural forest with samples from the trees planted in the artificial stands and 2) samples from the trees planted in the artificial stands with the natural regeneration in the artificial stands.

Significant differences in allele frequencies (nSSR) between pairs of samples were tested with a Fisher exact test (Raymond and Rousset, 1995) using the program Genepop 3.4 (available at http://wbiomed.curtin.edu.au/genepop/).

*2.4 Temporal change in allelic frequencies*

Allele frequencies change through time through the influences of gene flow and genetic drift (assuming neutral variation). Using the change in allele frequencies in samples of different generations has been mainly used for estimating effective population sizes (e.g. Waples, 1989). Recently Wang and Whitlock (2003) have developed a new maximum likelihood method that estimates immigration rates and

171 effective population sizes simultaneously. This method, implemented in the MLNE
172 program (available at http://www.zoo.cam.ac.uk/ioz/software.htm), has been used on
173 the nSSR data of the reforested samples using the planted trees as the parental
174 generation and the natural regeneration as the following generation. To estimate the
175 gene flow, the method also uses the allele frequencies of the potential external source
176 of genes, which in our case were taken from the surrounding natural stands.

177

178 *2.5 Additional analysis of gene flow*

179 A hierarchical AMOVA on cpSSRs was used to study the paternal heterogeneity from
180 the sampled seedlings, as in Dyer and Sork (2001). The chloroplast is paternally
181 inherited in conifers; therefore, cpSSR variation in the seedlings might reflect the
182 genetic diversity of the pollen clouds of the area. However, in our case, having
183 sampled seedlings on the ground, we cannot ignore some influence of seed dispersal
184 in our data. Canopy density can influence the dispersal of propagules diminishing
185 windspeed (see Nathan and Katul, 2005). Thus, sampling sites were nested into two
186 AMOVA groups, Arico (low vegetation cover) and Fasnia (high vegetation cover, del
187 Arco Aguilar et al., 1992).

188

189 The genotype assignment method developed by Rannala and Mountain (1997),
190 implemented in their program Immanc 5 (available at
191 http://www.rannala.org/labpages/software.html), was used on the nSSR data to detect
192 immigrants among the seedlings. This method allows performing the test for different
193 levels of immigrant ancestry considering the probabilities of external origin of: A) the
194 whole genotype (first generation immigrant, i.e. the individual has two foreign
195 parents), B) half of the genotype (second generation immigrant, i.e. the individual has one foreign parent), etc. The program calculates the relative probability for each individual to be born from parents with no recent immigrant ancestry rather than from both foreign parents (denoted $\Lambda_{d=0}$) or from one foreign parent (denoted $\Lambda_{d=1}$). These calculations are based on the allele frequencies of the studied population and the potential source populations using a Bayesian approach. The null distribution (individual with no immigrant ancestry) of the statistic $\ln \Lambda$ is built by a Monte Carlo simulation and $p$-values and statistical power (for $\alpha = 0.05$) are estimated for each test.

In our case the interpretation of the Rannala and Mountain (1997) assignment tests are slightly different, because we have effectively only one generation of possible immigrants. 'First generation' immigrants (whole genotype of external origin) will correspond to seeds developed and fertilised outside the stand. 'Second generation' immigrants (half genotype of external origin) will correspond to seeds from the stand fertilised with pollen from outside (the alternative possibility, seed from outside fertilised with pollen from the stand and dispersed into the stand, is considered highly improbable and will be ignored in the interpretation). The proportion of 'seed-immigrants' ($d = 0$) and 'pollen-immigrants' ($d = 1$) in the analysed seedlings will be estimates of seed and pollen immigration rates. To check the accuracy of the method we examined the results of these tests on the planted trees, which we consider by definition 'non-immigrants'. The proportion of planted trees detected as immigrants will give us the error rate of the method.

**3. Results and discussion**

*3.1 Genetic differences among samples*

220 Significant differentiation ($\chi^2$ = 50.438, d.f. = 16, $p$ < 0.001) in the nSSR allele
221 frequencies was detected between planted and natural forest. However theses
222 differences cannot be interpreted as introduction of foreign material as significant
223 differentiation ($\chi^2$ = 54.787, d.f. = 16, $p$ < 0.001) was also detected between the two
224 spatially close natural stands. Also, there were significant differences in the allele
225 frequencies between planted trees and understory natural regeneration for both
226 reforested stands (Arico: $\chi^2$ = 27.282, d.f. = 16, $p$ = 0.038 and Fasnia: $\chi^2$ = 28.005, d.f.
227 = 16, $p$ = 0.032). These changes may be the result of the effects of drift and gene flow
228 in the reforested stands (see next section).

229

230 Levels of nSSR genetic diversity in the reforested stands are, in general, similar to
231 those found in the surrounding natural forest (table 1, results for individual nSSR loci
232 not shown). Mean expected heterozygosity presented a significantly lower value in
233 the plantations but the diversity index based on allele number, which is more sensitive
234 to population bottlenecks (Nei et al., 1975), showed no significantly lower values and
235 in some cases allele diversity was even higher. Therefore, it is unclear what the lower
236 $H_e$ value indicates. The higher proportion of rare alleles (alleles with frequency <
237 0.01) in the reforested stands (27.45%) in comparison to the natural forest (19.42%)
238 suggests to us that a possible mixed origin of the seedling stock (or different seedling
239 origins for the different reforestation phases) could produce an accumulation of alleles
240 from different areas (explaining high levels of allelic diversity) but with lower
241 frequencies in the mixture (resulting in low $H_e$ values). In fact, observed
242 heterozygosity (mean $H_o$=0.648) is lower than the expected heterozygosity suggesting
243 the possibility of a Wahlund effect ($F_{IS}$=0.112, $p$-value = 0.006).

244

For cpSSRs, there is a significantly lower effective number of haplotypes in both artificial stands compared to natural forest (table 1). Bottlenecks associated with the reforestation process could have had a stronger effect on cpSSR diversity than in nSSRs because the effective population size for the chloroplast genome is half the nuclear effective population size. However, no significant reduction was found at $H_e$ and $D^2_{sh}$, probably because these measures are less sensitive to population size changes. Alternatively, the lower haplotype diversity found could be attributed to a low genetic diversity provenance for the seedlings used in the plantation. However, low diversity populations of *P. canariensis* are mainly marginal (Vaxevanidou et al. 2006) which are unlikely locations for seed collection by Forest Services.

In the comparison of the reforested areas with their natural regeneration we found a significant increase in the effective number of haplotypes and in the mean effective number of alleles (table 2). Although the increase on mean $A_e$ was not apparent when the two reforested stands were studied separately, the whole reforestation area had a significant gain in alleles for loci SPAC 11.8, PtTX3116, PtTX4001, ssrPt_ctg4698 and ssrPt_ctg7731 (data not shown). This increase of the allelic diversity in the reforestation may be attributable to gene flow from the natural stands.

*3.2 Gene flow into the artificial stands*

Three different methods have been used to explore the immigration dynamics within the reforested stands. First, we examined the temporal changes in allele frequencies between the planted trees and the following generation, represented by the understory natural regeneration. We applied the Wang and Whitlock (2003) maximum likelihood method using the allele frequencies in the natural forest as the potential external source of immigrants. The estimates (and 95% confidence intervals) of effective population size ($N_e$) and immigration rate ($m$) obtained from this method are $N_e$ = 36.88 (29.09–50.12) and $m$ = 0.68 (0.46–0.88) for Arico and $N_e$ = 28.83 (22.95–38.78) and $m$ = 0.75 (0.54–0.94) for Fasnia.

Two other methods were employed to characterize gene flow in the area but none of them shed further light on the subject. The results of the AMOVA test gave non-significant differentiation of the pollen diversity for both among sites and among stands (Φ-statistics < 0.033, $p$ > 0.17). This result suggests a high pollen dispersal capacity. A second method was the estimation of immigration rates by the identification of immigrants among the natural regenerated seedlings with the Rannala and Mountain (1997) assignment test. The immigration rates into the reforested patches, calculated as the proportion of identified immigrants over the sample size, are $m_{seed}$ = 0.07 and $m_{pollen}$ = 0.12. However, these estimates are unreliable as when the same assignment test was applied to the planted trees (considered 'non-immigrants') 15 individuals were identified as second generation immigrants (pollen dispersal) and 6 as first generation immigrants (seed dispersal). These error rates are in the same order of magnitude as the estimated immigration rates obtained. This poor performance of the test is due to the low genetic differentiation among the samples which resulted in very low statistical power (results not shown).

The high levels of gene flow suggested by these results are not surprising as pines are predominantly outcrossing and wind pollinated (Ledig, 1998). Previous estimates of immigration rates in artificial stands are available from pollen contamination studies in seed orchards. Among the factors influencing the pollen contamination levels in seed orchards are the distance to the nearest stands of the same species and the relative pollen production of the seed orchard to the surrounding forest (Adams and Burczyk, 2000). Wind direction can also increase pollen contamination locally (Yazdani and Lindgren, 1991). The range of pollen immigration rates found for pines in such studies are high, ranging from 0.26 to 0.75 (both estimates for *Pinus sylvestris* seed orchards, Harju and Muona, 1989; Yazdani and Lindgren, 1991) depending on the particulars of each case. Immigration rates calculated in natural stands from paternity analysis are also high 0.30–0.31 (minimum estimates, Lian et al., 2001; González-Martínez et al., 2003) but decrease to 0.05–0.07 in isolated populations (Schuster and Mitton, 2000; Robledo-Arnuncio and Gil, 2004). Therefore, the close proximity to big natural forest and the small size of the reforested stands studied offers some explanation for the high levels of immigration into the reforested area.

## 4. Conclusions

The effect of the reforestation process was detected in the chloroplast genome with a reduction in the effective number of haplotypes while the effects on the nuclear genome were uncertain. The difference in the effective population size between the two genomes could explain a different sensitivity to bottlenecks. These results suggest that, for the particular case studied, the planted stock originated from a reduced number of trees. Despite the lower haplotype diversity found within the plantations the richness of their future genetic pool does not seem jeopardized because both chloroplast and nuclear genetic diversities appear to increase within the understory natural regeneration. High levels of gene flow from the adjacent and larger natural stands explain this rise in diversity. These results highlight the importance of the presence and abundance of natural forest for the regeneration of genetically rich forest (see, for example, Glaubitz et al., 2003a; Glaubitz et al., 2003b). However, these conclusions should not be extended to other *P. canariensis* plantations, many of which are located far from natural stands (such as those found on north Tenerife, del Arco Aguilar et al., 1992). Current management of these reforested areas is mainly focused on the reduction of the number of trees to facilitate natural regeneration (Gil, 2006). This present work suggests the utility of studying the genetic makeup of artificial stands and their present or future natural regeneration. This may allow for the assessment of whether the reforestation process reduces genetic diversity (which will depend on the particular reforestation management done at different times and places) and, for those stands affected, their potential to recover genetic diversity in the following generations. This may identify stands for which the plantation of new individuals, from a controlled provenance and originating from numerous parental trees, may be beneficial for increasing their pool of genetic diversity.

**Acknowledgements**

We thank Cabildo Insular de Tenerife for sampling permit. University of East Anglia provided of a PhD scholarship to MN.

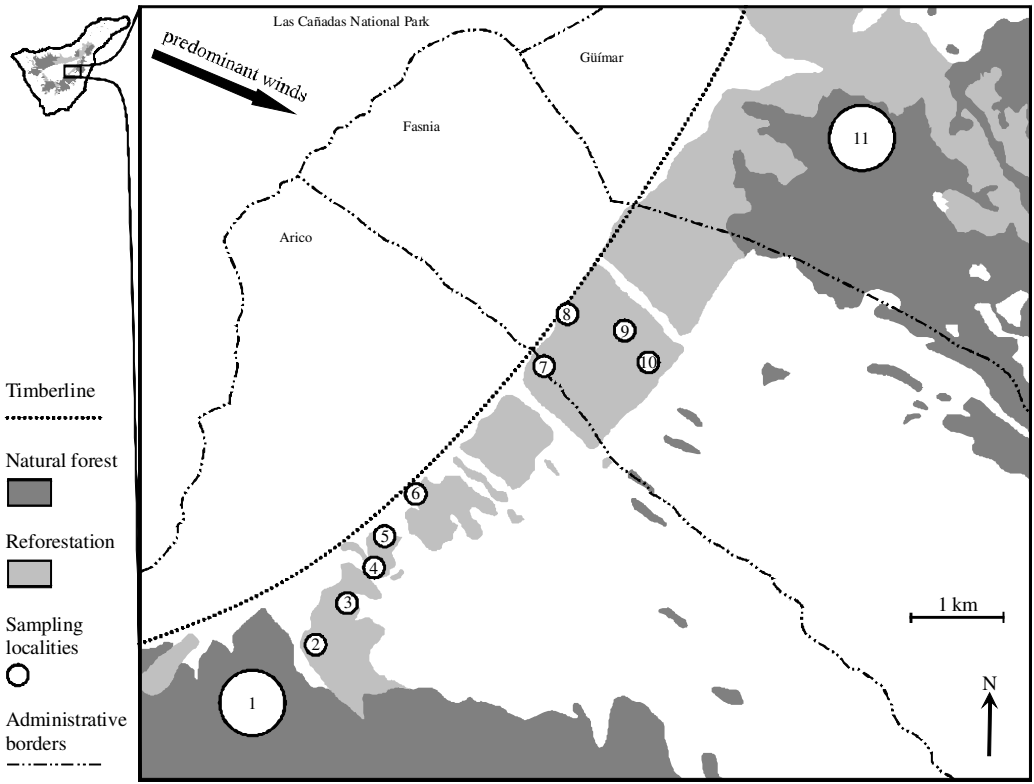

*Figure 1*.— Map of the study site on the South-eastern slope of Tenerife (Canary Islands). Sampling locations are marked with white circles proportional in diameter to the sampling size.

448 *Table 1.*— Diversity indices in the natural forest (pooled Arico and Güímar samples)

449 and reforested areas (with statistical significance of the difference with natural forest

450 in parentheses).

|  |  | Natural | Planted, both stands (*p*-value) | Arico, planted trees (*p*-value) | Fasnia, planted trees (*p*-value) |
|---|---|---|---|---|---|
|  | Sample size | 97 | 91 | 50 | 41 |
| chloroplast | $n_h$ | 49 | 46 (0.312) | 32 (< 0.001*) | 25 (< 0.001*) |
|  | $n_e$ | 34.496 | 19.051 (0.006*) | 26.085 (0.056) | 12.828 (< 0.001*) |
|  | $H_e$ | 0.971 | 0.948 (0.142) | 0.962 (0.287) | 0.923 (0.115) |
|  | $D^2_{sh}$ | 3.120 | 2.739 (0.261) | 2.584 (0.156) | 2.984 (0.461) |
| nuclear | mean $A$ | 17.735 | 19.125 (0.990) | 15.875 (0.895) | 13.875 (0.340) |
|  | mean $A_e$ | 9.618 | 9.335 (0.360) | 10.493 (0.832) | 8.583 (0.128) |
|  | mean $H_e$ | 0.753 | 0.729 (0.032*) | 0.726 (0.038*) | 0.733 (0.119) |

451 * Significant at α = 0.05

452

Table 2.— Diversity indices in the reforested areas and their understory regeneration, and statistical significance of the differences. Analysis done for the whole reforested area, reforested area in Arico (planted in 1981-85) and reforested area in Fasnia (planted in 1956-65). Sample sizes in parentheses.

| | | Planted (91) | Regeneration (90) | $p$-value |
|---|---|---|---|---|
| chloroplast | $n_h$ | 46 | 49 | 0.219 |
| chloroplast | $n_e$ | 19.051 | 35.768 | 0.006* |
| chloroplast | $H_e$ | 0.948 | 0.972 | 0.193 |
| chloroplast | $D^2_{sh}$ | 2.739 | 3.033 | 0.322 |
| nuclear | mean $A$ | 19.125 | 19.375 | 0.342 |
| nuclear | mean $A_e$ | 9.335 | 11.100 | 0.012* |
| nuclear | mean $H_e$ | 0.729 | 0.760 | 0.027* |
| | | Arico planted (50) | Arico regeneration (50) | $p$-value |
| chloroplast | $n_h$ | 32 | 36 | 0.106 |
| chloroplast | $n_e$ | 26.085 | 47.152 | 0.001* |
| chloroplast | $H_e$ | 0.962 | 0.980 | 0.211 |
| chloroplast | $D^2_{sh}$ | 2.584 | 3.194 | 0.163 |
| nuclear | mean $A$ | 15.875 | 14.250 | 0.949 |
| nuclear | mean $A_e$ | 10.493 | 10.511 | 0.489 |
| nuclear | mean $H_e$ | 0.726 | 0.738 | 0.213 |
| | | Fasnia planted (41) | Fasnia regeneration (40) | $p$-value |
| chloroplast | $n_h$ | 25 | 24 | 0.689 |
| chloroplast | $n_e$ | 12.828 | 22.314 | 0.037* |
| chloroplast | $H_e$ | 0.923 | 0.956 | 0.228 |
| chloroplast | $D^2_{sh}$ | 2.984 | 2.771 | 0.573 |
| nuclear | mean $A$ | 13.875 | 13.750 | 0.498 |
| nuclear | mean $A_e$ | 8.583 | 9.723 | 0.122 |
| nuclear | mean $H_e$ | 0.733 | 0.736 | 0.438 |

* Significant at $\alpha = 0.05$